\documentclass[12pt]{article}
\usepackage{amsmath}
\usepackage{graphicx,psfrag,epsf}
\usepackage{enumerate}
\usepackage{natbib}
\usepackage{url} 

\RequirePackage{amsthm,amsfonts,amssymb}
\usepackage{lineno}
\usepackage{array}
\usepackage{subfigure}
\usepackage{color}
\usepackage{enumitem}
\usepackage{cancel}
\usepackage{tikz}
\newtheorem{thm}{Theorem}

\newtheorem{cor}[thm]{Corollary}

\theoremstyle{definition}

\newcommand{\blind}{1}

\begin{document}

\def\spacingset#1{\renewcommand{\baselinestretch}%
{#1}\small\normalsize} \spacingset{1}


\if1\blind
{
  \title{\bf The Cost of Sequential Adaptation and the Lower Bound for Mean Squared Error}
  \author{Sergey Tarima\hspace{.2cm}\\
    Institute for Health and Equity, Medical College of Wisconsin\\
    and \\
    Nancy Flournoy \\
    Department of Statistics, University of Missouri}
  \maketitle
} \fi

\if0\blind
{
  \bigskip
  \bigskip
  \bigskip
  \begin{center}
    {\LARGE\bf The Cost of Sequential Adaptation}
\end{center}
  \medskip
} \fi

\bigskip
\begin{abstract}
Informative interim adaptations lead to random sample sizes.  The random sample size becomes a component of the sufficient statistic and estimation based solely on observed samples or on the likelihood function does not use all available statistical evidence.  The total Fisher Information (FI) is decomposed into the design FI and a conditional-on-design FI. The FI unspent by the interim adaptation is used to determine the lower mean squared error in post-adaptation estimation. Theoretical results are illustrated with simple normal samples collected according to a two-stage design with a possibility of early stopping.
\end{abstract}

\noindent%
{\it Keywords:}  Adaptive designs, adapted support, Cramer-Rao lower bound, group sequential designs, Fisher information, interim hypothesis testing
\vfill

\newpage
\spacingset{1.45} 

\section{Introduction} 
\label{Introduction}

Study designs with data-dependent sample sizes provide an attractive alternative to fixed sample size designs because interim decisions may reduce cost and decrease the duration of the study. Two groups of such sequential procedures, namely, groups sequential designs (GSD) and sample size re-estimation (SSR), are popular examples of studies with random sample sizes [see, for example, \citet{jennison1999, Proschan2006, chuang2006sample, friede2006sample}]. Our manuscript mostly deals with GSDs and SSR, but other informative adaptive designs may also benefit from the theoretical findings reported here. 

The benefits of interim adaptations come with certain costs as the distributions of sample-based statistics change [\citet{Armitage1969,efron1975, siegmund1985sequential, liu1999,liu2006,Tarima2021MPTest}]. Some recognized consequences are that  Bayesian procedures (that typically are not adjusted for multiple testing) do not control Type 1 error [see Table 18.1 in \citet{jennison1999}], maximum likelihood estimators become biased [\citet{whitehead1986bias}], and Wald confidence intervals do not provide the desired coverage after GSDs or SSRs.
In this paper we use FI to quantify the information loss associated with interim decisions using interim data that depends on the parameter of interest.

For simplicity of exposition, we focus on two-stage adaptive designs with a single interim analysis. Extension to multiple interim analyses follows from recursive application of the presented methods. For illustrative purposes we use a normal example with arbitrary sample sizes $n_1$ and $n_2$ for stage 1 and 2, respectively. 

Section \ref{sec:Notation} introduces notation in the context of two-stage designs without possibility of early stopping. Section \ref{sec:sequential_settings} shows the impact of the possibility of early stopping on support, sample space, and data distributions. In Section \ref{sec:Fisher_information}, FI is decomposed into the portion consumed by the interim decision [analogous to the Design Information in \citet{marschner2021general} that  strictly pertained to the Observed Information] and the remainder which  is represented as a weighted average of FI conditional on the possible interim decisions. Section \ref{sec:sim_nor_ex} illustrates results with numerical calculations for a simple normal example with $X_i \sim N(\theta,1)$, $i=1,2$ (that is, $ n_1=n_2=1$) under a design that has a possibility of early stopping after observing $X_1$ if $X_1>1.96$. Section \ref{Summary} concludes the manuscript with a short summary.

New results reported in Sections~\ref{sec:sequential_settings}, \ref{sec:Fisher_information} and \ref{sec:Cramer_Rao_lower_bounds} are  illustrated in Section \ref{sec:sim_nor_ex}.  Section \ref{sec:Notation} is necessary for understanding.

\section{Notation and Review of Basics for Independent Observations}\label{sec:Notation}

\spacingset{1} 
\begin{table}[bt!]
	\centering
	\begin{tabular}{|c|c|}
	\hline
		Feature & Description \\
		\hline
		r.v. & random variable \\
		$\theta$ & parameter of interest \\
		\hline
		$\mathbf{X}_d$      & stage-specific sample of r.v.s $\left(X_{n_{(d-1)}+1},\ldots,X_{n_{(d)}}\right)$\\
		$\mathbf{X}_{(d)}$  & cumulative sample of r.v.s  $\left(X_1,\ldots,X_{n_{(d)}}\right)$\\
		$\mathbf{x}_d$      & a realization of $\mathbf{X}_d$ \\
		$\mathbf{x}_{(d)}$      & a realization of $\mathbf{X}_{(d)}$\\
    \hline
		$n_{d}$  & stage-specific sample size\\
	    $n_{(d)}$  & $n_{(d)} = \sum_{k=1}^d n_{d}$ \\
		$t_{d}$  & $=\sum_{n_{(d-1)}+1}^{n_{(d)}} X_i$\\
	    $t_{(d)}$  & $=\sum_{1}^{n_{d}} X_i$ \\
		$Z_{d}$  & $=t_{d}/\sqrt{n_{d}}$\\
	    $Z_{(d)}$  & $=t_{(d)}/\sqrt{n_{(d)}}$ \\
    \hline
		$\hat\theta_{d}$  & stage-specific MLE $=t_{d}/n_{d}$\\
	    $\hat\theta_{(d)}$  & cumulative MLE $=t_{(d)}/n_{(d)}$\\
		$\tilde\theta_{d}$  & a stage-specific statistic $\tilde\theta_{d} = \tilde\theta_{d}\left(\mathbf{X}_d\right)$\\
	    $\tilde\theta_{(d)}$  & a cumulative statistic $\tilde\theta_{(d)} = \tilde\theta_{(d)}\left(\mathbf{X}_{(d)}\right)$\\
	    	\hline		
	${\cal{T}}$ & adapted-to-design support\\
	$\mathbf{X}_{\cal{T}}$ & $\mathbf{X} = \left(\mathbf{X}_1,\mathbf{X}_2\right)$ defined on ${\cal{T}}$\\
	\hline		
	 ${\cal{I}}_{\mathbf{X}_d}(\theta)$ & state-specific expected Fisher information\\
	 ${\cal{I}}_{\mathbf{X}_d = \mathbf{x}_d}^{obs}(\theta)$ & state-specific observed Fisher information\\
	 ${\cal{I}}_{\mathbf{X}_{\cal{T}}}(\theta)$ & cumulative expected Fisher information, or total if $d=2$\\
	 ${\cal{I}}_{\mathbf{X}_{\cal{T}}|D}(\theta)$ & conditional-on-design cumulative expected Fisher information\\
	 ${\cal{I}}_{\mathbf{X}_{\cal{T}}|D=d}(\theta)$ & conditional-on-a-realized-design expected Fisher information\\
	\hline
	\end{tabular}
	\caption{Notation: decisions $d=1,2$; $n_0=0$; $\prod_{i=1}^0[\cdot] = 1$.}
	\label{Definitions}
\end{table}

\spacingset{1.45}
In this section, we assume that there is no impact of interim analyses and the data are aggregated into two samples only for convenience.  A summary of our notation is given in Table~\ref{Definitions}.

Let $\mathbf{X}_1 = \left(X_1,\ldots,X_{n_1}\right)$ and $\mathbf{X}_2 = \left(X_{n_1+1},\ldots,X_{n_1+n_2}\right)$ be two samples of independent random variables with $X_i\sim f_{X}\left(x|\theta\right)$. Joint $d$-stage-specific densities $f_{\mathbf{X}_d}\left(\mathbf{X}_d|\theta\right)$ are used to define the FI in $\mathbf{X}_d$, $d\in\{1,2\}$: $${\cal{I}}_{\mathbf{X}_d}(\theta) =Var\left[\frac{\partial}{\partial \theta} \log f_{\mathbf{X}_d}(\mathbf{x}|\theta)\right].$$ If $\mathbf{X}_d=\mathbf{x}_d$ alone is observed, we assume that the log-likelihood function $l(\theta|\mathbf{X}_d=\mathbf{x}_d)$ has a unique maximum  at $\widehat\theta_d$.  If both $\mathbf{X}_1=\mathbf{x}_1$ and $\mathbf{X}_2=\mathbf{x}_2$ are  observed independently, $$l(\theta|\mathbf{X}_1=\mathbf{x}_1, \mathbf{X}_2=\mathbf{x}_2) = l(\theta|\mathbf{X}_1=\mathbf{x}_1) + l(\theta|\mathbf{X}_2=\mathbf{x}_2)$$  denotes their log-likelihood function and its  maximum is denoted $\widehat\theta_{(2)}$. Note that we use parentheses to distinguish estimators based on cumulative data (e.g., $\hat\theta_{(2)}$ uses data from both stages 1 and 2) from stage-specific estimators (e.g., $\hat\theta_{2}$ is based on stage 2 data only).

\subsection{Cramer-Rao Lower Bound (CRLB) if no early stopping is possible} \label{CRLB_basics} Let $\tilde\theta_d = \tilde\theta_d\left(\mathbf{X}_d\right)$ be a stage-specific estimator for which $\textrm{E}[\tilde\theta_d]=\theta + b_d(\theta)$ and $\left(\partial/\partial \theta\right) f_{\mathbf{X}_d}(t|\theta)$ exist and $\partial/\partial \theta$ can be passed under the integral sign in $\int f_{\mathbf{X}_d}(t|\theta) dt = 1$ and $ \int t \ f_{\mathbf{X}_d}(t|\theta) dt$. Then 
 \begin{equation}
     \textrm{Var}[\tilde\theta_d] \ge \frac{\left[1+\frac{\partial}{\partial \theta}b_d(\theta)\right]^2}{{\cal{I}}_{\mathbf{X}_d}(\theta)}
 \end{equation}
 and
  \begin{equation}
     \textrm{E}\left[(\tilde\theta_d-\theta)^2\right] \ge \frac{\left[1+\frac{\partial}{\partial \theta}b_d(\theta)\right]^2}{{\cal{I}}_{\mathbf{X}_d}(\theta)} + b_d^2(\theta).
 \end{equation}
 In a similar manner, for an estimator $\tilde\theta_{(2)}$ defined  $\mathbf{X}_{(2)}=\left(\mathbf{X}_1,\mathbf{X}_2\right)$:
 \begin{equation}\label{eq:CRLB0}
     \textrm{Var}[\tilde\theta_{(2)}] \ge \frac{\left[1+\frac{\partial}{\partial \theta}b_{(2)}(\theta)\right]^2}{{\cal{I}}_{\mathbf{X}_{(2)}}(\theta)}
 \end{equation}
 and
  \begin{equation}\label{eq:MSELB0}
     \textrm{E}\left[(\tilde\theta_{(2)}-\theta)^2\right] \ge \frac{\left[1+\frac{\partial}{\partial \theta}b_{(2)}(\theta)\right]^2}{{\cal{I}}_{\mathbf{X}_{(d)}}(\theta)} + b^2_{(2)}(\theta),
 \end{equation}
 where $\textrm{E}[\tilde\theta_{(2)}]=\theta + b_{(2)}(\theta)$.
 If $\mathbf{X}_1$ and $\mathbf{X}_2$ are independent  ${\cal{I}}_{\mathbf{X}_{(2)}} = {\cal{I}}_{\mathbf{X}_1} + {\cal{I}}_{\mathbf{X}_2}$. 
 
 The Cramer-Rao lower bound (CRLB) was suggested independently by  \citet{rao1945information} and \citet{cramer1946contribution} under the assumption of fixed sample sizes.
 

\subsection{Normal Example} Let $t_d=\sum_{i=n_{(d-1)}+1}^{n_{(d)}} x_i$. Now if $X_i\sim N(\theta, 1)$, the  stage-specific MLE is $\hat\theta_d = n_d^{-1}t_d \sim N(\theta,n_d^{-1})$ 
with FI  $${\cal{I}}_{\mathbf{X}_d}(\theta)
={\cal{I}}^{obs}_{\mathbf{X}_d=\mathbf{x}_d}(\theta)
 = -\left(\partial^2/\partial \theta^2\right) l(\theta|\mathbf{X}_d=\mathbf{x}_d)=n_d.$$
The random variable $(\hat\theta_1,\hat\theta_2)$ is defined on the probability space $\left(R^2,{\cal{B}}, P\right)$, where $R^2$ is the sample space [$R=\left(-\infty,\infty\right)$]; 
${\cal{B}}$ is the Borel $\sigma$-algebra on $R^2$; and 
 if both $\mathbf{X}_1=\mathbf{x}_1$ and $\mathbf{X}_2=\mathbf{x}_2$ are 
 observed independently, the probability measure $P$ is a bivariate normal distribution with mean $(\theta,\theta)$, variances $n_1^{-1}$ and $n_2^{-1}$, and zero correlation.  Then, the observed and expected Fisher informations are equal and 
 $${\cal{I}}^{obs}_{\mathbf{X}_1=\mathbf{x}_1, \mathbf{X}_2=\mathbf{x}_2}(\theta) =  -\frac{\partial^2}{\partial \theta^2}l(\theta|\mathbf{X}_1=\mathbf{x}_1)=-\frac{\partial^2}{\partial \theta^2}l(\theta| \mathbf{X}_2=\mathbf{x}_2)=n_1+n_2.$$
 
\section{Two-Stage Experiments with the Possibility of Early Stopping} \label{sec:sequential_settings} Let $Z_1= Z_1\left(\mathbf{X}_1\right)$ be an interim test statistic based on stage one data alone. What is the distribution of two stage-specific statistics $\tilde\theta_1=\tilde\theta_1\left(\mathbf{X}_1\right)$ and $\tilde\theta_2=\tilde\theta_2\left(\mathbf{X}_2\right)$ in \textit{sequential settings} when $\tilde\theta_2$ is only observed if $z_1 < c_1$, where $c_1$ is some pre-determined critical value? 

Let $D$ denote an interim rule for choosing a decision from a decision space ${\cal{D}}$. In GSDs with one interim analysis, the interim decision rule is \begin{equation}\label{GSD.decisions}
{\cal{D}} = \begin{cases}
1=\text{`stop'} & \text{if } z_1 \ge c_1;\\ 2=\text{`collect }n_2\text{ extra observations'}  & \text{if } z_1 < c_1.
\end{cases}\end{equation} 

\subsection{The Probability Distribution of the Observable Random Variables} \label{prob_dist} In such sequential settings, the joint support of $(\tilde\theta_1,\tilde\theta_2$) changes from that of independent statistics as $\tilde\theta_2$ is not just missing but impossible when $Z_1 \ge c_1$. \citet{Tarima2021MPTest} observe that because some sampling combinations of $\tilde\theta_1$ and $\tilde\theta_2$ are not observable, 
the support of $\tilde\theta_1$ and $\tilde\theta_2$ becomes $${\cal{T}}=\left\{\tilde\theta_1 | z_1 \ge c_1 \right\} \cup \left\{ \{\tilde\theta_1 | z_1 < c_1\} \cap \{\tilde\theta_2 \in R\} \right\}.$$ 
Note that ${\cal{T}} \subset R^2$. If $A \in {\cal{T}}$,  then the probability measure on the $\sigma$-algebra, $\sigma({\cal{T}})$ is
\begin{align}
    \mu_{{\cal{T}}}=\text{Pr}[(\tilde\theta_1, \tilde\theta_2) \in A] &= \notag 
\text{Pr}(Z_1 \ge c_1)\text{Pr}(\tilde\theta_1  \in A|Z_1 \ge c_1) \\ &\qquad \qquad + \notag
\text{Pr}(Z_1 < c_1)\text{Pr}[(\tilde\theta_1, \tilde\theta_2)  \in A|Z_1 < c_1].
\end{align} 
On the new probability space $\{{\cal{T}}, \sigma({\cal{T}}), \mu_{{\cal{T}}}\}$, we have an observable random variable, namely,
\begin{align*}
 \tilde\theta = \begin{cases}
\tilde\theta_1 &\textrm{ if } $D=1$ \\ (\tilde\theta_1,\tilde\theta_2) & \textrm{ if }  $D=2$.
\end{cases}
\end{align*}
Factor the density of ${\tilde\theta_1}$ as \begin{equation} \label{eq:density1}
    f_{\tilde\theta_1}(y_1) = \sum_{d=1}^2f_{D}(d) [f_{\tilde\theta_1|D}(y_1|d)]^{I(D=d)}= f_{D}(d) [f_{\tilde\theta_1|D}(y_1|d)]^{I(D=d)},
\end{equation}  where
 $f_D(d) = \prod_{d=1}^2[\text{Pr}(D=d)]^{I(D=d)}$  is the probability distribution  of the Bernoulli random variable $D$ and
 $f_{\tilde\theta_1|D=d}$ is a \textit{random}  probability distribution for different realizations of $D$: \begin{align}\label{eq:5}f_{\tilde\theta_1|D=d}(y_1|d) = f_{\tilde\theta_1}(y_1)\frac{I(D=d)}{\text{Pr}(D=d)} = \frac{\left[f_{\tilde\theta_1}(y_1)\right]^{I(D=d)}}{\text{Pr}(D=d)}.
 \end{align} 
Alternatively, $\tilde\theta_1$ can be viewed as a mixture of $\tilde\theta_1|D=1$ and $\tilde\theta_1|D=2$ with \begin{equation}
f_{\tilde\theta_1}(y_1) = \text{Pr}(D=1)f_{\tilde\theta_1|D=1}(y_1|D=1) + \text{Pr}(D=2)f_{\tilde\theta_1|D=2}(y_1|D=2).\label{decompZ1}
\end{equation}
The same decomposition applied to $f_{\tilde\theta}(\mathbf{y})$ with $\mathbf{y}=(y_1,y_2)$ provides
 \begin{equation}
f_{\tilde\theta}(\mathbf{y}) = \text{Pr}(D=1)f_{\tilde\theta_1|D=1}(y_1|D=1) + \text{Pr}(D=2)f_{\left(\tilde\theta_1,\tilde\theta_2\right)|D=2}(\mathbf{y}|D=2).\label{decompZT}
\end{equation}
Thus, $f_{\tilde\theta_1}(y_1)$ and  $f_{\tilde\theta}(\mathbf{y})$ both can be described as mixtures of distributions. 

Using $f_{\tilde\theta_1|D=1}(y_1|D=1)$ from Equation (\ref{eq:5}) and the analogous expression 
\begin{align}
f_{\tilde\theta|D=2}(\mathbf{y}|D=2) = f_{\left(\tilde\theta_1,\tilde\theta_2\right)}(\mathbf{y})\frac{I(D=2)}{\text{Pr}(D=2)} = \frac{\left[f_{\left(\tilde\theta_1,\tilde\theta_2\right)}(\mathbf{y})\right]^{I(D=2)}}{\text{Pr}(D=2)}
 \end{align} 
 in Equation (\ref{decompZT}), another useful representation for the density $f_{\tilde\theta}(\mathbf{y})$ is 
\begin{align}
f_{\tilde\theta}(\mathbf{y}) &=\left[f_{\tilde\theta_1}(y_1)\right]^{I(D=1)} + \left[f_{\left(\tilde\theta_1,\tilde\theta_2\right)}(\mathbf{y})\right]^{I(D=2)}\nonumber \\ &=\left[f_{\tilde\theta_1}(y_1)\right]^{I(D=1)}  \left[f_{\left(\tilde\theta_1,\tilde\theta_2\right)}(\mathbf{y})\right]^{I(D=2)},\label{two_subdens}
\end{align}
where the right-hand side of the mixture density can be represented as a sum or as a product of two \textit{sub}densities.

\subsection{The Likelihood Function's Insensitivity to  Early Stopping Possibilities}
The log-likelihood function is conditional on  $\mathbf{X}_1=\mathbf{x}_1$, and on  $\mathbf{X}_2=\mathbf{x}_2$ if stage-two observations are made, and consequently on the stopping stage $d$:
\begin{align*}
 l(\theta|\mathbf{X}_1=\mathbf{x}_1, \mathbf{X}_2=\mathbf{x}_2, D=d) =  \begin{cases}
l(\theta|\mathbf{X}_1=\mathbf{x}_1) &\textrm{ if } D=1, \\ l(\theta|\mathbf{X}_1=\mathbf{x}_1, \mathbf{X}_2=\mathbf{x}_2) & \textrm{ if }  D=2.
\end{cases}
\end{align*}
Importantly, the log-likelihood function is the same as in the case of independent observations without the possibility of early stopping. More generally, the likelihood is insensitive to the effect of early stopping  mechanisms.  This insensitivity is inherited by the score function, the MLE, and the observed Fisher information. 

The expected Fisher information, however, changes with early stopping options [see Section \ref{sec:Fisher_information}]. Consider three  experiments with $X_i \sim N(\theta,1)$: \begin{description}
    \item[$E_1:$] An experiment with no early stopping having a fixed sample size $n_1+n_2$ and a joint sampling density $\prod_{i=1}^{n_1+n_2}f_{X_i}(x_i|\theta)$. 
    \item[$E_2:$] A sequential experiment with early stopping if $\sqrt{n_1}\bar X_1 > 1.96$ [motivated by a one-stage experiment with $\alpha=0.025$ and no planned interim test] has joint sampling density $\prod_{i=1}^{n_1}f_{X_i}(x_i|\theta)  \left[\prod_{i=n_1+1}^{n_1+n_2}f_{X_i}(x_i|\theta)\right]^{I\left(\sqrt{n_1}\bar X_1 \textcolor{blue}{<} 1.96\right)}$.
    \item[$E_3:$] A sequential experiment with  early stopping if $\sqrt{n_1}\bar X_1 > 2.78$ [motivated by  the \citet{o1979multiple} stopping boundary] has  joint sampling density\\ $\prod_{i=1}^{n_1}f_{X_i}(x_i|\theta)  \left[\prod_{i=n_1+1}^{n_1+n_2}f_{X_i}(x_i|\theta)\right]^{I\left(\sqrt{n_1}\bar X_1 \textcolor{blue}{<} 2.78\right)}$.
\end{description}
The sampling density in $E_1$ is defined on ${R}^2$, whereas the densities under $E_2$ and $E_3$ have support $\mathbf{\cal{T}}$ as described in Section \ref{prob_dist}.  

Suppose each of the three experiments has been run, and $E_2$ and $E_3$ did not stop early because $\sqrt{n_1}\bar X_1 > 1.96$ and $\sqrt{n_1}\bar X_1 > 2.78$ are observed, respectively. Suppose further that  same data (outcomes) are observed in all three experiments: $\mathbf{X}_1=\mathbf{x}_1$
    and $\mathbf{X}_2=\mathbf{x}_2$.  Then despite different underlying designs and densities, all experiments ($E_1$, $E_2$ and $E_3$) share the same likelihood function: 
\begin{equation}
L(\theta|X_1=x_1,\ldots,X_{n_1+n_2}=x_{n_1+n_2})=\prod_{i=1}^{n_1+n_2}f_{X_i}(x_i|\theta). 
\end{equation} 
The likelihood $L(\theta|X_1,\ldots,X_{n_1+n_2})$ is a random variable, but its generating distributions differ across the three experiments, which illustrates that $L(\theta|X_1,\ldots,X_{n_1+n_2})$ is insensitive to the study design.
This is the same phenomenon as the well-known proportionality of the binomial and negative binomial likelihoods  [see Example 9 in \citet{BergerWolpert} and Example 6.3.7 on page 295 in \citet{casella}].


\subsection{Normal Example (Two-Stage Design, Arbitrary $n_1$ and $n_2$)} The log-likelihood function is
\begin{align*}
 l\left(\theta\Big|\hat\theta_{(d)}=\frac{t_{(d)}}{n_{(d)}}, D=d\right) =  
 \frac{1}{2\pi}
\exp\left(-\frac{1}{2}n_{(d)}\left(t_{(d)}-\theta\right)^2\right), 
\end{align*}
where $t_{(d)}=\sum_{i=1}^{n_{(d)}}x_i$, and the value of $\hat\theta_{(2)}$ is not observed when $D=2$. The interim test statistic is $Z_1=t_1/\sqrt{n_1}$ and the MLE  
\begin{align}
\widehat\theta &= I(z_1 \ge c_1)\frac{t_1}{n_1} + I(z_1 < c_1)\left(\frac{t_{(2)}}{n_{(2)}}\right) \notag \\
&= I(z_1 \ge c_1)\hat\theta_1 + I(z_1 < c_1)\left(\frac{n_1}{n_{(2)} }\hat\theta_1 + 
\frac{n_2}{n_{(2)}} \hat\theta_2\right) \notag \\
&= I(z_1 \ge c_1)\hat\theta_1 + I(z_1 < c_1)\hat\theta_{(2)},   
\end{align}
is a random variable defined on $\left({\cal{T}}, \sigma({\cal{T}}), \mu_{{\cal{T}}}\right)$. Significantly, $\widehat\theta$ is a mixture of a left truncated normal random variable $\{\hat\theta_1|D=1\}$ and the weighted average of a right truncated normal  $\{\hat\theta_1|D=2\}$ and a normal $\hat\theta_2$. Thus, in this sequential setting, $\widehat\theta$ is no longer a normal random variable.

\section{Fisher Information with Informative Interim Decisions}
\label{sec:Fisher_information}
To this point we have focused on simple binary interim decisions (Section \ref{FI.GSD}), but to accommodate SSR procedures, Section \ref{FI.SSR} introduces a wider decision space and more complex decision functions;  FI is decomposed to expose the proportion consumed by the interim decision options for both GSD and SSR procedures.

\subsection{Fisher Information with an Interim Test} \label{FI.GSD}
Assuming standard regularity conditions [e.g. \citet{Ferguson1996}], FI in $\mathbf{X}_1$ conditional on the interim decision variable $D$ is the variance of the score function for 
$f_{\mathbf{X}_1|D}(y_1|D)$:
\begin{align}
    {\cal{I}}_{\mathbf{X}_1|D}(\theta) &=\textrm{Var} \left[\frac{\partial}{\partial \theta}\log f_{\mathbf{X}_1|D}(y_1|D)\right]\notag \\&= \int_{-\infty}^{\infty} -\frac{\partial^2}{\partial \theta^2} [\log f_{\mathbf{X}_1|D}(y_1|D)] f_{\mathbf{X}_1}(\mathbf{x}_1) d\mathbf{x}_1. \label{cond_on_D} 
\end{align}
In contrast, FI in $\mathbf{X}_1$ conditional on a specific \textit{realization}~$d$ of an interim decision is
\begin{align}
\label{cond_on_realization}
&{\cal{I}}_{\mathbf{X}_1|D=d}(\theta) = \int_{\infty}^{\infty} I(D=d) \frac{\partial^2}{\partial \theta^2} \left[-\log f_{\mathbf{X}_1|D}(\mathbf{x}_1|d)\right] f_{\mathbf{X}_1|D}(\mathbf{x}_1|d)d\mathbf{x}_1 \notag \\ 
&\hspace{-6pt} = \begin{cases}\displaystyle
 \int_{-\infty}^{\infty}  
 \frac{\partial^2}{\partial \theta^2} \left[-\log \frac{f_{\mathbf{X}_1}(\mathbf{x}_1)}{\text{Pr}(z_1 \ge c_1)} \right] \frac{f_{\mathbf{X}_1}(\mathbf{x}_1)I(z_1 \ge c_1)}{\text{Pr}(z_1 \ge c_1)}d\mathbf{x}_1 &\hspace{-8pt} \text{ if } z_1 \ge c_1; \\[10pt]
\displaystyle
\int_{-\infty}^{\infty}  \frac{\partial^2}{\partial \theta^2} \left[-\log \frac{f_{\mathbf{X}_1}(\mathbf{x}_1) }{\text{Pr}(z_1 < c_1)}\right] \frac{f_{\mathbf{X}_1}(\mathbf{x}_1)I(z_1 < c_1)}{\text{Pr}(z_1 < c_1)}d\mathbf{x}_1 & \hspace{-8pt}\text{ if } z_1 < c_1.
\end{cases}
\end{align}
We note that the  integration is multiple to highlight the fact that $\mathbf{x}_1$ has $n_1$ components and each is integrated out.

The information measures defined by Equations \eqref{cond_on_D} and \eqref{cond_on_realization} differ by the type of conditioning: ${\cal{I}}_{\mathbf{X}_1|D}(\theta)$  averages over distributions of $\mathbf{X}_1$ (and $D$ is fully determined by $\mathbf{X}_1$), whereas  ${\cal{I}}_{\mathbf{X}_1|D=d}(\theta)$ averages over the distribution of the \textit{conditional} random variable $\mathbf{X}_1|D=d$. Conditioning on a random variable, as in (\ref{cond_on_D}), is used in \cite{zegers2015fisher} for example, while conditioning on an observed constant, as in (\ref{cond_on_realization}), is used in \cite{Mihoc2003}. 

While in this manuscript, we condition is on $D=d$ unless stated otherwise.     However, ${\cal{I}}_{\mathbf{X}_1|D}(\theta)$ provides an important measure of the information left in $\mathbf{X}_1$ after informative adaptation. 
Taking the $\log$ of the joint density $f_{\mathbf{X}_1}(\mathbf{x}_1) =  f_{D}(d)  f_{\mathbf{X}_1|D} (\mathbf{x}_1|d)^{I(D=d)}$
and using the law of total variance on the score function, FI  in $\mathbf{X}_1$ and
$\mathbf{X}_{\cal{T}}$, respectively, can be written as
\begin{eqnarray}
{\cal{I}}_{\mathbf{X}_1}(\theta) &=& {\cal{I}}_{D}(\theta) + {\cal{I}}_{\mathbf{X}_1|D}(\theta) \label{infZ1_1}\nonumber \\
&=& {\cal{I}}_{D}(\theta) + \text{Pr}(D=1|\theta){\cal{I}}_{\mathbf{X}_1|D=1}(\theta) +
\text{Pr}(D=2|\theta){\cal{I}}_{\mathbf{X}_1|D=2}(\theta) \label{infZ1}
\end{eqnarray}
and
\begin{align} \label{info_partitioning}
{\cal{I}}_{\mathbf{X}_{\cal{T}}}(\theta) =& {\cal{I}}_{D}(\theta) + \text{Pr}\left(D=1|\theta\right) {\cal{I}}_{\mathbf{X}_1|D=1}(\theta) + \text{Pr}\left(D=2|\theta\right) \left[{\cal{I}}_{\mathbf{X}_1|D=2}(\theta) + {\cal{I}}_{\mathbf{X}_2}(\theta) \right] \notag  \\ =&
{\cal{I}}_{D}(\theta) + {\cal{I}}_{\mathbf{X}|D}(\theta) \notag  \\ =& {\cal{I}}_{D}(\theta) + {\cal{I}}_{\mathbf{X}_1}(\theta) + \text{Pr}\left(D=2|\theta\right)  {\cal{I}}_{\mathbf{X}_2}(\theta),
\end{align} 
where $ {\cal{I}}_{D}(\theta)$ is
the portion of the total FI that  is consumed by permitting one interim test in a two-stage design (the design information): 
\begin{equation}
    {\cal{I}}_{D}(\theta) = \textrm{Var}\left[\frac{\partial}{\partial\theta}\log f_{D}(d)\right]=\sum_{d=1}^2 \frac{\partial^2}{\partial \theta^2} \left[-\log \text{Pr}(D=d|\theta)\right] \text{Pr}(D=d|\theta).
\end{equation}

\subsection{Fisher Information with Sample Size Re-estimation} \label{FI.SSR}
Given a sample size re-estimation formula, there are still many ways to construct the procedure to be used in an application.  For example, if the calculated estimate is below the current sample size $n_1$, the procedure may call for the study to be stopped; if the estimate exceeds a given limit it will be truncated.  To accommodate such practical considerations, we define the decision rule as a function of the available data $\mathbf{x}_1$ rather than the summary $Z$-statistic. 

Let $\left\{{\cal{C}}_1,\ldots,{\cal{C}}_K\right\}$ be a partition of all possible realizations of stage 1 data. Then the SSR can be formalized by the following decision function
\begin{equation}\label{interim.decisions}
{\cal{D}} = \begin{cases}
1=\text{`stop'} & \text{if } \mathbf{x}_1  \in {\cal{C}}_1\\ 2=\text{`collect }n_2^{(2)}\text{ extra observations'}  & \text{if } \mathbf{x}_1 \in {\cal{C}}_2\\
\vdots\ =\, \vdots &\vdots\\
K=\text{`collect }n_2^{(K)}\text{ extra observations'}  & \text{if } \mathbf{x}_1 \in {\cal{C}}_K
\end{cases}
\end{equation}
The decision space specified by \eqref{interim.decisions} accommodates all possible sample size recalculation procedures (e.g., procedures based on the observed variability of the treatment effect,  the observed treatment effect itself, etc.).  The resulting support $\left({\cal{T}}\right)$ and the probability measure of the observable random variables are defined as generalizations of those detailed in Section~\ref{prob_dist}.

Note that  different ``labels'' may be assigned to the $\{1,\ldots, K\}$ decisions in Equation \eqref{interim.decisions}  to describe other types of interim decisions  (e.g., joint futility and efficacy stopping, enrichment procedures). 

The sequential decomposition of FI with an interim test shown in Equation (\ref{info_partitioning}) is extended  to
designs that have multiple decision options at the time of the interim analysis in Theorem~\ref{inf_decomp}:
\begin{thm}\label{inf_decomp} If a decision $d \in \{1,\ldots,K\}$ can be made at an interim analysis, the following decomposition of Fisher Information applies:
\begin{align} \label{info_partitioning_mult}
{\cal{I}}_{\mathbf{X}_{\cal{T}}}(\theta) =& {\cal{I}}_{D}(\theta) + \sum_{d=1}^K\text{Pr}\left(D=d|\theta\right) \left[{\cal{I}}_{\mathbf{X}_1|D=d}(\theta) + {\cal{I}}_{\mathbf{X}_2|D=d}(\theta)\right].\end{align} 
\end{thm}
When the interim decision $D=d$ is made, the amount of FI available for further inferential (e.g., further hypothesis testing and estimation) consists of  the previous information conditional on $D=d$ [${\cal{I}}_{\mathbf{X}_1|D=d}(\theta)$] and  the new information to be collected before next analysis [${\cal{I}}_{\mathbf{X}_2|D=d}(\theta)$]. 
 
 If there is more than one interim analysis, Theorem \ref{inf_decomp} can be applied sequentially with $\mathbf{X}_1$ referring to the data collected before an interim analysis and $\mathbf{X}_2$ referring to the data that might be collected after the interim analysis.  Thus, Theorem \ref{inf_decomp} describes how Fisher Information is allocated across the whole decision tree branching overtime with interim analyses.

 Section \ref{sec:sim_nor_ex} presents a simple normal example to illustrate the theoretical findings.

\section{Lower Bound for Mean Squared Error in Sequential Experiments}
\label{sec:Cramer_Rao_lower_bounds}
Section \ref{CRLB_basics} reviews how Fisher information appears in the CRLB inequality that determines the minimum variance, and more generally the minimum MSE among all regular estimators. Moreover, this bound is reached in one-parameter exponential family models where the MLE absorbs all statistical evidence about the canonical parameter $\theta$. Section \ref{sec:Fisher_information} shows that the  situation changes with informative interim adaptations: not all Fisher information is available for post-adaptation inference. Thus, CRLB inequality needs to be extended to account for interim adaptations.

\begin{thm}{\textbf{(Lower Bounds for Mean Squared Error  in Sequential Experiments)}}\label{th:CRLB}
If, for any interim decision $d \in \{1,\ldots, K\}$, ${\cal{I}}_{\mathbf{X}_{\cal{T}}|D=d}(\theta)$ exists for any $\theta$ and $\frac{\partial}{\partial \theta}E(\tilde\theta\mid D=d)=E(\frac{\partial}{\partial \theta}\tilde\theta\mid D=d)$, then
\begin{enumerate}
\item conditionally on $D=d$, the lower boundary for the MSE of $\tilde\theta$ is 
  \begin{equation} \label{eq:CRLB.d}
     \textrm{E}\left([\tilde\theta_{(d)}-\theta]^2\right) \ge \frac{\left[1+\frac{\partial}{\partial \theta}b(\theta)\right]^2}{{\cal{I}}_{\mathbf{X}_{\cal{T}}|D=d}(\theta)} + b_{(d)}^2(\theta)
 \end{equation}
\item and, unconditionally, the lower bound is 
  \begin{equation}  \label{eq:CRLB}
     \textrm{E}\left([\tilde\theta-\theta]^2\right) \ge \sum_{d=1}^K P_d(\theta)\left[ \frac{\left[1+\frac{\partial}{\partial \theta}b_{(d)}(\theta)\right]^2}{{\cal{I}}_{\mathbf{X}_{\cal{T}}|D=d}(\theta)} + b_{(d)}^2(\theta)\right],
 \end{equation}
 \end{enumerate}
where $E(\tilde\theta_{(d)}) = \theta + b_{(d)}(\theta)$. 
\end{thm}

\textbf{Proof of Theorem \ref{th:CRLB}:} Existence of Fisher information and interchangeability of integration and differentiation ensures existence of the Cramer-Rao lower boundary at each $D=d$, which proves inequality  (\ref{eq:CRLB.d}). From  $\textrm{E}([\hat\theta-\theta]^2)=\sum_{d=1}^K P_d(\theta) \textrm{E}([\hat\theta_{(d)}-\theta]^2)$, inequality (\ref{eq:CRLB}) immediately follows.
\textbf{Q.E.D.}

\begin{cor}{\textbf{The MLE for a canonical parameter in the one-parameter exponential family attain the CRLB}.}
\label{cor:CRLB.exp}
The MLE $\hat\theta$ in the one-parameter exponential family  reaches the CRLB with minimum MSE:
 \begin{equation}
     \textrm{E}\left([\hat\theta-\theta]^2\right)=\sum_{d=1}^K P_d(\theta) \left(\frac{\left[1+\frac{\partial}{\partial \theta}b(\theta)\right]^2}{{\cal{I}}_{\mathbf{X}_{\cal{T}}|D=d}(\theta)} + b_{(d)}^2(\theta)\right).
 \end{equation}\end{cor}
\textbf{Proof of Corollary \ref{cor:CRLB.exp}:} \citet{Tarima2021MPTest} showed that the possibility of early stopping after an MLE-based interim hypothesis
test of the form $\hat\theta_1 > c_1$
changes the distribution of the test statistic, but conditionally on $D=d$, $\hat\theta_{(d)}$ continues being sufficient and to belong to the exponential family. Same argument applies for interim decision specific $\hat\theta_{(d)}$, where interim decision are defined by (\ref{interim.decisions}). Thus, under Theorem \ref{th:CRLB} assumptions, the Cramer-Rao lower boundary is reached for any $D=d$:
 \begin{equation}
     \textrm{E}\left([\hat\theta_{(d)}-\theta]^2\right) = \frac{\left[1+\frac{\partial}{\partial \theta}b(\theta)\right]^2}{{\cal{I}}_{\mathbf{X}_{\cal{T}}|D=d}(\theta)} + b_{(d)}^2(\theta)
 \end{equation}
 and ${\cal{I}}_{\hat\theta_{(d)}}(\theta) = {\cal{I}}_{\mathbf{X}_{\cal{T}}|D=d}(\theta)$. Further, from $$\textrm{E}\left([\hat\theta-\theta]^2\right)=\sum_{d=1}^K P_d(\theta) \textrm{E}\left([\hat\theta_{(d)}-\theta]^2\right)$$ the result follows.
 \textbf{Q.E.D.}

 Section \ref{sec:sim_nor_ex} presents a simple normal example to illustrate the theoretical findings.

\section{Normal Example, $n_1=n_2=1$, $X_i \sim N(\theta,1)$\label{sec:sim_nor_ex}} 
If $n_1=n_2=1$, $\mathbf{X}_1=X_1$, $\mathbf{X}_2=X_2$, and the MLE $\hat\theta=X_1$ when $D=1$ and $\hat\theta=(X_1+X_2)/2$ when $D=2$. The density of $\hat\theta$ if stopped early $(D=1)$ is $$f_{\hat\theta|D=1}(t_1)=\frac{\phi(t_1-\theta)}{\Phi(c_1-\theta)};$$
if the experiment proceeds to stage 2 ($D=2$), the density is the convolution 
\begin{align}\label{dens2}
f_{\hat\theta|D=2}\left(t_{(2)}\right)&=\frac{1}{1-\Phi(c_1-\theta)}\int_{-\infty}^{c_1} \left(\frac{1}{\sqrt{2\pi 0.5}}\right)^2 \nonumber \\
&\times \exp\left[-\frac{1}{2}\left(\frac{t_{(2)}-t_1/2-\theta/2}{1/2}\right)^2\right] \exp\left[-\frac{1}{2}\left(\frac{t_1/2-\theta/2}{1/2}\right)^2\right]dt_1 \nonumber \\
&=\frac{1}{1-\Phi(c_1-\theta)}\int_{-\infty}^{c_1}\frac{1}{\pi} \phi\left(2t_{(2)}-t_1-\theta\right) \phi\left(t_1-\theta\right)dt_1.\end{align}
Minor algebra shows that Fisher information in a normal random variable $X_1 \sim N(\theta,1)$ truncated to the interval $[a,b]$ is equal to its variance
\begin{equation}\label{tr_inf}
    1-A^{-1}\left[b_{st}\phi(b_{st}) - a_{at}\phi(a_{st})\right] - 
    A^{-2}\left[\phi(b_{st})-\phi(a_{st})\right]^2,
\end{equation}
where $b_{st} = b-\theta$, $a_{st} = a-\theta$ and $A=\Phi(b_{st})-\Phi(a_{st})$.

\spacingset{1} 
\begin{table}
\begin{center}
\begin{tabular}{l|c|l} 
 \hline
 Information & Value & Comment \\[2pt] 
 \hline
 \multicolumn{3}{l}{Informative Stopping, $c_1=1.96,\; \text{Pr}(D=1\mid\theta)=0.5,\ \theta=1.96$}\\[2pt] 
 \hline
 ${\cal{I}}_{X_1|D=1}$ & 0.3634 & \\ 
 ${\cal{I}}_{X_1|D=2}$ & 0.3634 & \\ 
 ${\cal{I}}_{X_1|D}$   & 0.3634 & $=0.5\cdot 0.3634-0.5\cdot 0.3634$\\ 
 ${\cal{I}}_{D}$       & 0.6366 & $=1-0.3634$ \\ 
 ${\cal{I}}_{\mathbf{X}_{\cal{T}}|D=2}$ & 1.3634 & =1+0.3634 \\ 
  ${\cal{I}}_{\mathbf{X}_{\cal{T}}|D}$ & 0.8634 & $=0.5 \cdot 0.3634 + 0.5 (0.3634 + 1)$ \\ 
 ${\cal{I}}_{\mathbf{X}_{\cal{T}}}$ & 1.5 & $=0.6366 + 0.5 \cdot 0.3634 + 0.5 (0.3634 + 1)$ \\ 
\hline
\multicolumn{3}{l}{Informative Stopping, $c_1=1.96,\;\text{Pr}(D=1\mid\theta)=0.025,\;\theta=0$}\\[2pt] 
 \hline
 ${\cal{I}}_{X_1|D=1}$ & 0.8789 & \\ 
 ${\cal{I}}_{X_1|D=2}$ & 0.3634 & \\ 
 ${\cal{I}}_{X_1|D}$   & 0.8598 & $=0.025 \cdot 0.1167 + 0.975\cdot 0.8789$\\ 
 ${\cal{I}}_{D}$       & 0.1402 & $=1-0.8598$ \\ 
 ${\cal{I}}_{\mathbf{X}_{\cal{T}}|D=2}$ & 1.8789 & =1+0.8789 \\ 
  ${\cal{I}}_{\mathbf{X}_{\cal{T}}|D}$ & 1.8349 & $=0.025 \cdot 0.1167 + 0.975 (0.8789 + 1)$ \\ 
 ${\cal{I}}_{\mathbf{X}_{\cal{T}}}$ & 1.975 & $=0.1402 + 0.025 \cdot 0.1167 + 0.975 (0.8789 + 1)$ \\ 
\hline
\multicolumn{3}{l}{Non-Informative Stopping with $\text{Pr}(D=1|\theta)=0.5$, $\forall \theta$}\\[2pt]
 \hline
 ${\cal{I}}_{X_1|D=1}$ & 1 & \\ 
 ${\cal{I}}_{X_1|D=2}$ & 1 & \\ 
 ${\cal{I}}_{X_1|D}$   & 1 & \\ 
 ${\cal{I}}_{D}$       & 0 & \\ 
 ${\cal{I}}_{\mathbf{X}_{\cal{T}}|D=2}$ & 2 & $=1+1$ \\ 
  ${\cal{I}}_{\mathbf{X}_{\cal{T}}|D}$ & 1.5 & $=1+0.5(1+1)-0$  \\ 
 ${\cal{I}}_{\mathbf{X}_{\cal{T}}}$ & 1.5 & $=1+0.5(1+1)$\\ 
\hline
\multicolumn{3}{l}{Non-Informative Stopping with $\text{Pr}(D=1|\theta)=0.025$, $\forall \theta$}\\[2pt] 
 \hline
 ${\cal{I}}_{X_1|D=1}$ & 1 & \\ 
 ${\cal{I}}_{X_1|D=2}$ & 1 & \\ 
 ${\cal{I}}_{X_1|D}$   & 1 & \\ 
 ${\cal{I}}_{D}$       & 0 & \\ 
 ${\cal{I}}_{\mathbf{X}_{\cal{T}}|D=2}$ & 2 & $=1+1$ \\ 
  ${\cal{I}}_{\mathbf{X}_{\cal{T}}|D}$ & 1.975 & $=1+0.975(1+1)-0$  \\ 
 ${\cal{I}}_{\mathbf{X}_{\cal{T}}}$ & 1.975 & $=1+0.975(1+1)$\\ 
\hline
\end{tabular}
\caption{Numerically calculated  values of various information measures given $n_1=n_2=1$, $X_i \sim N(\theta,1)$\label{tab1}}
\end{center}
\end{table}
\spacingset{1.45} 

Formula (\ref{tr_inf}) permits calculation of ${\cal{I}}_{\mathbf{X}_1|D=1}(\theta)$ and ${\cal{I}}_{\mathbf{X}_1|D=2}(\theta)$ in the 
Fisher information partitioning equation
(\ref{info_partitioning}). Other components (\ref{info_partitioning}) are easily calculable using untruncated normal formulas: ${\cal{I}}_{\mathbf{X}_1}(\theta) = {\cal{I}}_{\mathbf{X}_2}(\theta) = 1$; and the design information simplifies as
\begin{align*}
 {\cal{I}}_{D} &= E_{X_1}\left(-\frac{\partial^2}{\partial \theta^2}\log \text{Pr}(X_1 > c_1)\right) = E_{X_1}\left(-\frac{\partial^2}{\partial \theta^2}\log \Phi(c_1-\theta)\right)\\
& = 1 - \text{Pr}(D=1){\cal{I}}_{\mathbf{X}_1|D=1}(\theta)-\text{Pr}(D=2) {\cal{I}}_{\mathbf{X}_1|D=2}(\theta).
\end{align*}

\begin{figure}
\centering
\begin{subfigure}{{(a) $c_1=1.96$}}\label{fig:Ia}
\includegraphics[width=0.6\textwidth]{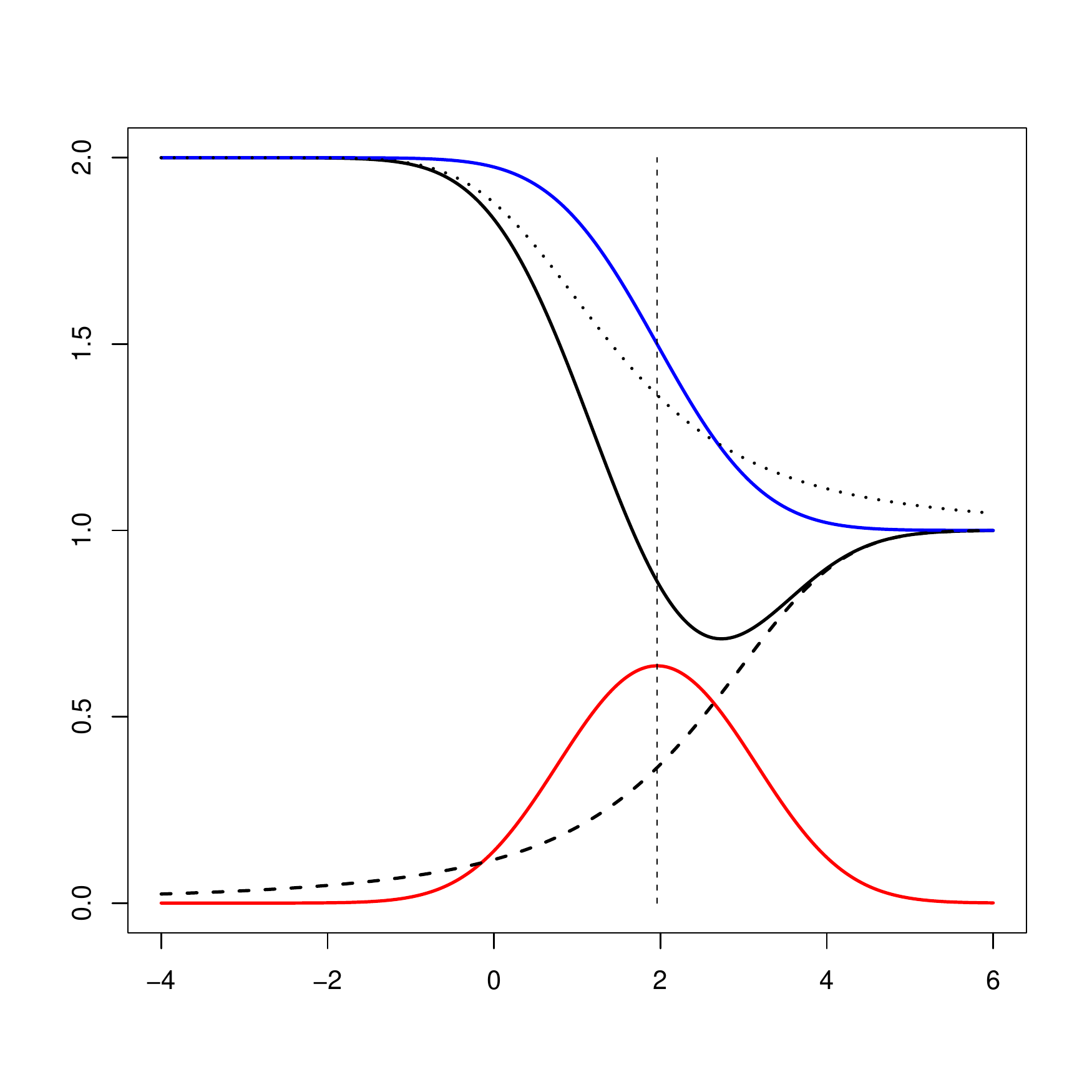}
\end{subfigure}\\
\begin{subfigure}{{(b) $c_1=2.78$}}\label{fig:Ib}
\includegraphics[width=0.6\textwidth]{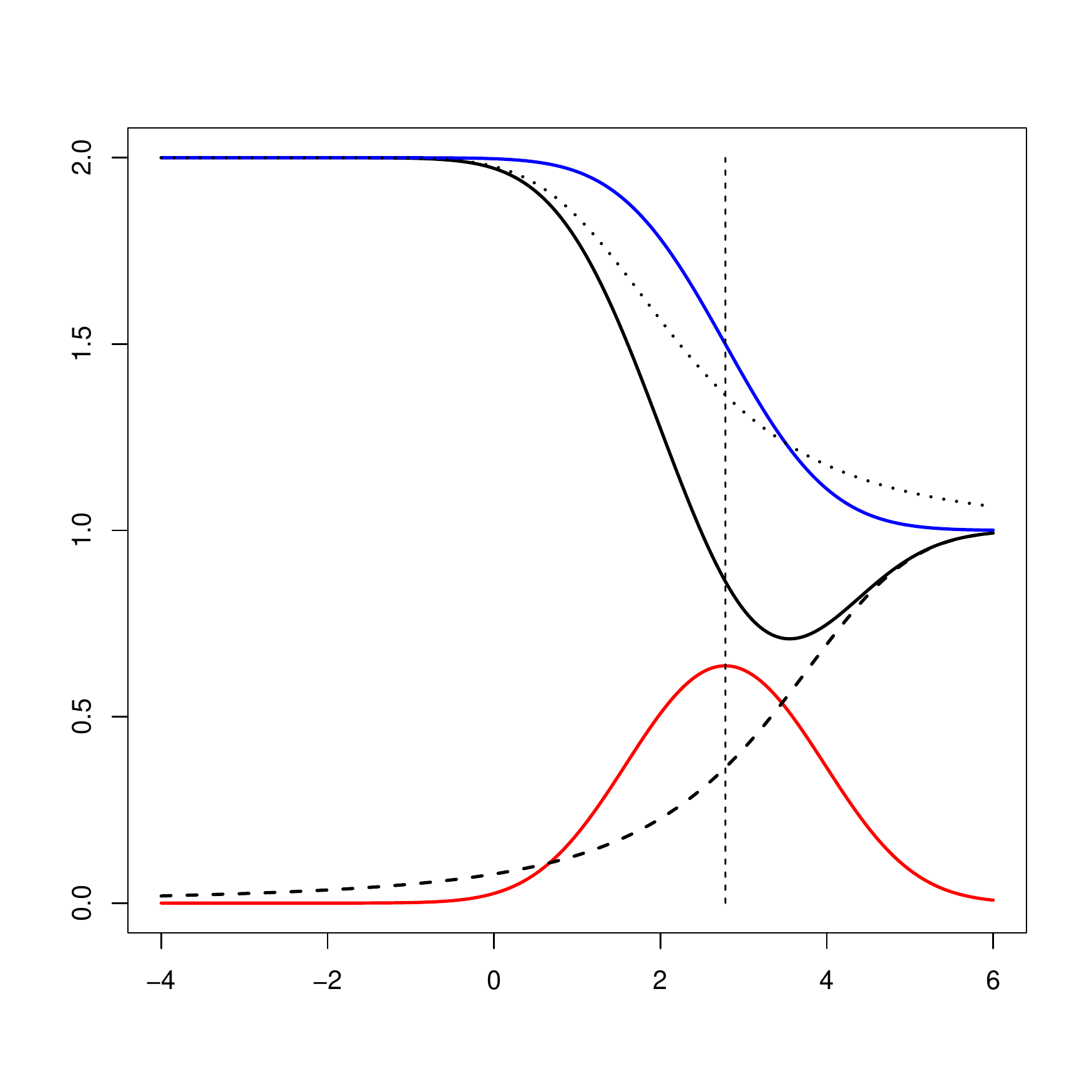}
\end{subfigure}
\caption{Fisher Information (FI) with $n_1=n_2=1$ versus $\theta$;  total FI in the sequential experiment is shown by a solid blue line, FI conditional on design - solid black, FI in design - solid red, FI conditional on $D=1$ - dashed black and FI conditional on $D=2$  - dotted black; is shown by a thin dashed vertical line}
\label{Fisher.Info} 
\end{figure}

Various measures of information are plotted in Figure~\ref{Fisher.Info} as a function of $\theta$. The two subfigures shown are constructed under  experiments E2 (an interim test that stops the experiment if $z_1\ge c_1=1.96$) and E3 (an interim test that stops the experiment if $z_1\ge c_1=2.78$).  
The total FI in an experiment with two independent normal random variables without an interim stopping option is 2; this is the maximum information available with stopping options. 
The total Fisher information (FI) with the interim stopping options (the blue curves) is $1+1\cdot \text{Pr}\left(z_1\ge c_1\mid \theta\right)$; it decreases from the maximum 2.0 for small values of $\theta$ to 1.0 for large $\theta$, reflecting the changing probability of observing one or two observations. 
If $\theta$ equals the critical value, then the chance of continuing is equal to the chance of stopping and the total FI is 1.5.

The total FI is  seen to decompose into the information conditional on the design (solid black) and the information in the design (solid red).  The design possibility of an interim test  consumes the greatest amount of information when $\theta$ equals the critical value [$\theta=1.96$ in panel (a) and $\theta=2.78$ in panel (b)] , and  tends to zero as $\theta\to\pm \infty$. Panels (a) and (b) illustrate that different stopping rules cause the same likelihood  to be associated with different information measures.

The information curves conditional on stopping at $D=1$ (dashed black line) increases from zero to 1.0, while the information conditional on stopping at $D=2$ (dotted black line) decreases from 2.0 to 1.0.

For more concrete comparison, Table \ref{tab1} reports the actual numerical calculations of ${\cal{I}}_{X_1|D=1}$,  ${\cal{I}}_{X_1|D=2}$, ${\cal{I}}_{X_1|D}$, ${\cal{I}}_{D}$, ${\cal{I}}_{\mathbf{X}_{\cal{T}}|D=2}$, ${\cal{I}}_{\mathbf{X}_{\cal{T}}|D}$, and ${\cal{I}}_{\mathbf{X}_{\cal{T}}}$ for two interim decision rules (informative stopping with $c_1=1.96$ and non-informative stopping) under $\theta=0$ and $\theta=1.96$. 

Table \ref{tab:sim_nor2} provides numerical calculations of the lowest possible mean squared error for the design with early stopping defined by $c_1=1.96$ under $\theta=0$ and 1.96. Per Corollary \ref{cor:CRLB.exp}, this lower bound for MSE is reached by the MLE. The first and third number in each set were verified by simulation.  Note that  if one makes the ``right''  interim decision, bias and mean squared error are reduced.  That is if the experiment stops earlier and the true $\theta$ is higher, the bias and mean squared error for post-testing estimation are reduced; if the experiment proceeds to the second stage and $\theta$ is lower, the bias and mean squared error are reduced as well.

\begin{table}
\begin{center}
\begin{tabular}{l|c|l} 
 \hline
 Information & Value & Comment \\ 
 \hline
 \multicolumn{3}{l}{Informative Stopping, $c_1=1.96,\;\theta=1.96,\;\phi(0)=0.3989, \;\Phi(0)=0.5$}\\
 \hline
 $E\left(\hat\theta-\theta|D=1\right)$ & 0.7979 & $=0.3989 \cdot 2$ \\ 
 $\frac{\partial}{\partial\theta}E\left(\hat\theta-\theta|D=1\right)$ & -0.6366 & $=-4 \cdot 0.3989^2$ \\ 
 $E\left[\left(\hat\theta-\theta\right)^2|D=1\right]$  & 1 & $=(1-0.6366)^2/0.3634+0.7979^2$\\ 
 \hline
 $E\left(\hat\theta-\theta|D=2\right)$ & -0.3989 & \\ 
 $\frac{\partial}{\partial\theta}E\left(\hat\theta-\theta|D=2\right)$ & -0.3183 & $=-4 \cdot 0.3989^2/2$ \\ 
 $E\left[\left(\hat\theta-\theta\right)^2|D=2\right]$  & 0.5 & $=(1-0.3183)^2/1.3634+(-0.3989)^2$\\ 
 \hline
\multicolumn{3}{l}{Informative Stopping, $c_1=1.96,\;\theta=0,\;\phi(1.96)=0.0584, \;\Phi(0)=0.975$}\\
 \hline
 $E\left(\hat\theta-\theta|D=1\right)$ & 2.3378 & $=0.0584/(1-0.9750)$ \\ 
 $\frac{\partial}{\partial\theta}E\left(\hat\theta-\theta|D=1\right)$ & -0.8833 & $=(1.96 \cdot 0.0584 \cdot 0.0250 - 0.0584^2)/(0.0250^2)$\\ 
 $E\left[\left(\hat\theta-\theta\right)^2|D=1\right]$  & 5.5821 & $=(1-0.8833)^2/0.1167 + 2.3378^2$\\ 
 \hline
 $E\left(\hat\theta-\theta|D=2\right)$ & -0.0300 & $=-0.0584 /(2 \cdot 0.9750)$ \\ 
 $\frac{\partial}{\partial\theta}E\left(\hat\theta-\theta|D=2\right)$ & -0.0605 & $=-0.5 \cdot (1.96 \cdot 0.0584 \cdot 0.9750 + 0.0584^2)/(0.9750^2)$ \\
 $E\left[\left(\hat\theta-\theta\right)^2|D=2\right]$  & 0.4706 & $=(1-0.0605)^2/1.8789+(-0.0300)^2$\\ 
 \hline
\end{tabular}
\caption{Bias conditional on interim decision $D$ for $\theta=1.96$ and $\theta=0$, its derivative, and the mean squared error based on analytic formulas at $n_1=n_2=1$, $X_i \sim N(\theta,1)$\label{tab:sim_nor2}}
\end{center}
\end{table}

\section{Summary}\label{Summary}
The cost of  adaptation is quantified.  We show how the cost of an interim test for early stopping varies with the parameter of interest and how it changes with the specific stopping rule even though  the likelihood is insensitive to the study design and the data collection procedure. 


Theorem \ref{inf_decomp} decomposes Fisher Information into a component resulting from the inclusion of informative decisions in the design, and separate components for before and after  interim decision points. Formula (\ref{info_partitioning}) details this decomposition for an experiment having a single interim test. This decomposition of the total Fisher information allows us to make the following useful observations:
\begin{itemize}
     \item With non-informative adaptations, ${\cal{I}}_{D}(\theta) = 0$; no information is lost for likelihood-based inference. For example, because the variance estimate of a normal random variable is independent of the estimated effect size, there is no loss of information  when sample size re-estimation is based only on an estimate of the  standard deviation  at the time of the interim analysis.
     \item It is possible that all the information in stage 1 is used for sample size re-estimation, in which case none is left for final estimation.
      \item The higher ${\cal{I}}_{D}(\theta)$ is the less Fisher information is left for estimation after stopping. 
     \item The  Fisher Information available in design, ${\cal{I}}_{D}(\theta)$, is a measure of the cost of informative stopping. This information is not absorbed by the likelihood function and is lost for likelihood-based inference. More generally, any estimators based solely on the observed data do not use the design information. 
    \item With one interim test for early stopping, the greatest information loss occurs when the critical value $c_1$ is equal to the parameter of interest $\theta$. At this value of $\theta$  post-testing estimation is affected the most.
    \item Fisher information available for estimation of $\theta$ or further testing after stopping at stages 1 or proceeding to stage 2 is  ${\cal{I}}_{\mathbf{X}_1|D=1}(\theta)$ and ${\cal{I}}_{\mathbf{X}_1|D=2}(\theta) + {\cal{I}}_{\mathbf{X}_2}(\theta)$, respectively.
\end{itemize}

Since some Fisher Information is spent for interim adaptation, the amount available for post-adaptation inference is smaller than the total. This is one reason why Cramer-Rao lower bound needs to be adjusted for informative interim adaptations. Another reason is the change in support of observable random variables. A sequential version of the Cramer-Rao lower bound suggested in \citet{wolfowitz1947efficiency} does not apply to designs with informative interim adaptations: see \citet{SIMONS198067} for  examples illustrating when Wolfowitz's bound does not work.

Theorem \ref{th:CRLB} gives a new bound for the smallest mean squared error, and as shown in Corollary \ref{cor:CRLB.exp} this lower bound is reached by the MLE in one-parameter exponential family with canonical parameterization.

We anticipate that our results will be useful for deriving and justifying optimal sequential designs with a dual goal of sequential testing and post-test estimation.

\section*{Conflict of interest}
 The authors declare that they have no conflict of interest.

\section*{Data Availability}
Data sharing is not applicable to this article as no new data were created or analyzed in this study. 

\section*{Funding} 
This research received no specific grant from any funding agency in the public, commercial, or not-for-profit sectors.

\bibliographystyle{Chicago}
\bibliography{bib}

\spacingset{1} 
\spacingset{1.45}

\end{document}